\documentclass[prb,twocolumn,showpacs,preprintnumbers,amsmath,amssymb,superscriptaddress]{revtex4}

\usepackage{graphicx}
\usepackage{dcolumn}
\usepackage{color}
\usepackage{multirow}

\allowdisplaybreaks
\newcommand{\beq}{\begin{equation}}
\newcommand{\eeq}{\end{equation}}
\newcommand{\beqa}{\begin{eqnarray}}
\newcommand{\eeqa}{\end{eqnarray}}

\begin{document}
\title{Tuning the work function in transition metal oxides and their heterostructures}

\author{Z.~Zhong}
\affiliation{Max-Planck-Institut f\"ur Festk\"orperforschung,
  Heisenbergstrasse 1, 70569 Stuttgart, Germany}

\author{P.~Hansmann}
\affiliation{Max-Planck-Institut f\"ur Festk\"orperforschung,
  Heisenbergstrasse 1, 70569 Stuttgart, Germany}

\begin{abstract}
The development of novel functional materials in experimental labs combined
with computer-based compound simulation brings the vision of materials design
on a microscopic scale continuously closer to reality. For many applications
interface and surface phenomena rather than bulk properties are key. One of
the most fundamental qualities of a material-vacuum interface is the energy
required to transfer an electron across this boundary, i.e. the work function. It is a crucial parameter for numerous applications, including organic electronics, field
electron emitters, and thermionic energy converters. Being generally very
resistant to degradation at high temperatures, transition metal oxides present
a promising materials class for such devices. We have performed a systematic
study for perovskite oxides that provides reference values and,
equally important, reports on materials trends and the \emph{tunability} of
work functions. Our results identify and classify dependencies of the work
function on several parameters including specific surface termination, surface
reconstructions, oxygen vacancies, and heterostructuring.
\end{abstract}
\date{\today}
\maketitle

\section{Introduction}
\label{Intro}
One of the main goals of computational solid state physics is the simulation
of ``hypothetical'' compounds that have not been synthesized  in the
experimental lab. Driven by its remarkable predictive power, Density
functional theory (DFT) is today the most important tool in the field. Together with the high level of sophistication that synthesis
technology has reached, the vision of materials design (i.e. the composition of
functional materials that are tuned for usage in specific devices) seems to
become reality.  
Control on the atomic level in materials synthesis e.g. with modern molecular
beam epitaxy and pulsed laser deposition lead to an increasing focus on heterogeneous superstructures and
effects associated with interfaces and surfaces. Especially oxide
superstructures \cite{Mannhart:sci10, Triscone:review2011} attracted lots of attention due to
partially extraordinary physics\cite{Ohtomo:nat04, Mannhart:SCLAOSTO} unknown in the bulk
materials but also adatom lattices or graphene grown on functional substrates
are in the focus of current experimental and theoretical studies.

In our DFT study we concentrate on a particular quality
of functional materials which is largely affected by its surface and crucial
to many applications: the work function $\Phi$. Devices that make use of thermionic electron
emission\cite{Yamamoto:rpp06}, catalytic surface properties\cite{Suntivich:Natc11}, construction of Schottky barriers\cite{Hikita:apl07,
  Minohara:prb10,Yajima:natc15}, or the conception of organic electronics
\cite{Greiner:natm12, Greiner:NPGAsia13} are some of the technologies for which knowledge of the work function is
essential. One of the main motivations for our study can be attributed to the very recent conception of so called thermoelectronic
devices \cite{Meir:JRSE13} which rely on the thermionic emission of
an emitter and the subsequent condensation on a collector material. Being, so
to speak, the next evolutionary step following thermionic energy convertors,
the new devices strive for a breakthrough in thermal to electrical energy
conversion. Two main aspects are key for the novel setups: i) stability
towards surface degradation also at elevated temperatures and ii) emitter and
collector materials with work functions tuned to one another. Due to these two
criteria we focus our study on transition metal oxide (TMO) materials: Most TMOs 
are thermally very stable and have high melting points. Moreover, we know from an extensive body of 
research that TMOs are sensitive to external perturbations (i.e. they are \emph{tunable}) which lead to rich phase 
diagrams\cite{Imada:rmp98}. 

If we turn to past studies of density functional theory on materials work
functions we find a good amount of research for simple metals
\cite{Skriver:prb92, Methfessel:prb92,Singh-Miller:prb09}, molecular
structures\cite{Rusu:prb06, deBoer:ADVMAT05} and simple oxides like MgO and ZnO
\cite{Giordano:prb05,Giordano:prl08} on metal surfaces,  Sc$_2$O$_3$ with adsorbed Ba
\cite{Jacobs:jpcc14}, modified silicon (111) surfaces \cite{Arefi:PhysChem14}, and even
graphene \cite{Giovannetti:PRL08, Young-JunYu:NANOLett09}. 
Yet, TMO work functions have been rarely studied and only recently started to attract
attention\cite{Suchitra:jap14, Kumar:jap14}.

In the present study we clarify the sensitivity of TMO work functions with respect to the specific
surface termination, surface relaxation, surface reconstruction, defect
structures (i.e. oxygen vacancies), externally induced surface strain,
electronic interactions on the mean-field level (i.e. inclusion of a Hubbard
U), and most importantly, material trends for a number of perovskite oxides and
superstructures. Our findings will not only serve as a reference for the
presented compounds but especially the observed parameter trends present a first
systematic step towards a broader understanding of how to push a compounds work function to the desired value.

The manuscript is organized in the following way: After reporting details of
our calculation scheme in section \ref{methods} we divide our results in three
sections. In section \ref{resultsA} we report on the sensitivity of the work
function on ``external'' and calculation parameters. While some of the
calculation parameters serve purely as a DFT benchmark  (e.g. choice of the
DFT functional or the Hubbard interaction U), others will be quite relevant
for comparison to experimental dependencies (e.g. lattice strain or oxygen
vacancies). In section \ref{resultsB} we explore the material
trends for different ABO$_3$ perovskite oxides. In section
\ref{resultsC} we consider the potential of tuning work functions with
heterostructuring oxide materials.

\section{Background and calculation details}
\label{methods}
The work function is defined as the minimal energy required to remove an
electron from inside the material across its surface into the vacuum. 
Conceptually the work function can be devided into a ``bulk'' and ``surface''
dependent part. If there were no charge redistribution at a materials
surface and the vacuum potential would be set to V$_{\rm vac.}=0$, the work function would be
equivalent to $\Phi=V_{\rm vac.}-E_F$ (where $E_F$ is the
materials Fermi energy).
(see e.g. Ref.~\onlinecite{Ashcroft_and_Mermin}). 
In reality, however, ionic and electronic charge in the vicinity to the
surface is very different from the
bulk and an additional electric field is generated by the non vanishing dipole
moment of the shifted charge arrangement. It is intuitively clear that
generally such a field, and therefore also the work function, depends on the
specific surface indices and termination. For our studies we
assume clean surfaces with well defined terminations. Let us remark already
here that while there are many materials, in particular simple metals
\cite{Michaelson:jap77}, which show little dependence of the work function on
microscopic details of the surface, TMO work functions are extremely sensitive
to these details.  

\begin{figure}[t!]
  \includegraphics[width=0.5\textwidth]{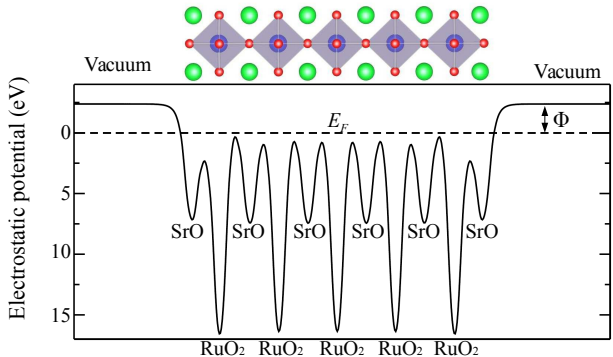}
  \caption{\label{Fig1} 
    Plane averaged electrostatic potential of a symmetric SrRuO$_3$ slab
    consisting of six SrO layers, five RuO$_2$  layers, and a 20 \AA vacuum. The
    electrostatic potential is defined with respect to Fermi energy $E_F$, and
    converges to a constant value in the vacuum region that is the work function
    $\Phi$  of SrRuO$_3$ with SrO surface termination. $\Phi$  therefore indicates
    the required energy to remove an electron at Fermi level from the material to
    a state at rest in the vacuum nearby the surface.}
\end{figure}

In this study we have selected a number of perovskite transition metal oxides
ABO$_3$  (where A=Ca, Sr, Ba; B=3d, Ti-Co; 4$d$: Zr-Rh) and their
heterostructures. The ABO$_3$ structure can be viewed as simple cubic lattice
of A atoms with a body centered B atom and oxygen atoms in the face centers.  
In the following we consider surfaces along the (001) direction which are the
most commonly studied surfaces in thin film or heterostructure compounds. In
this direction the crystal is build up by an alternating stacking of AO and
BO$_2$ layers \citep{Ohtomo:nat04}. The (001) ABO$_3$ then have either AO or
BO$_2$ surface terminations, and thus two intrinsically different work
functions. For the calculation of the work function within a DFT framework we
employ a symmetric slab geometry which is sketched in the top part of
Fig.~\ref{Fig1} for the example of SrRuO$_3$.  The SrO terminated SrRuO$_3$
slab consists of six SrO layers, five RuO$_2$  layers, and a vacuum thickness
of 20\AA; for BO$_2$ terminated surface (not shown in Fig.~\ref{Fig1}) we add
an additional BO$_2$ layer on each side of the slab. The work function $\Phi$
is then calculated as the energy difference between the plane averaged
electrostatic potential (excluding the exchange correlation part) of the slab
in the vacuum region and the Fermi level as can be seen in Fig.~\ref{Fig1};
Benchmark calculations confirmed that $\Phi$ is converged with this setup. 
For further information on the calculation of the averaged electrostatic
potential we refer to \onlinecite{Giovannetti:PRL08} and \onlinecite{Rusu:prb06}.

Since we are mostly interested in materials and superstructures grown on
substrates, the in-plane lattice constant was fixed to a=3.905\AA  which is that
of an assumed undistorted cubic SrTiO$_3$ substrate (it is also 
very close to the bulk lattice constant a=3.923\AA   of cubic SrRuO$_3$
\cite{Samata2009623}). For all calculations the internal atomic positions were
relaxed. The calculations were performed with the VASP (Vienna ab initio
simulation package) code \citep{Kresse:prb99} using the generalized gradient
approximation GGA -PBE functional \citep{PerdewPRL96} for electronic exchange
and correlation and a 16$\times$16$\times$1 k-point grid including the
$\Gamma$ point. In a set of selected benchmark calculations we compare the GGA
results also to those obtained by a local-density approximation (LDA)
functional\citep{Perdew:prb81}. While the latter one generally leads to
somewhat larger values of the work function, our main conclusions about materials trends
and sensitivities remain unchanged by the choice of the functional. 
Let us also explicitly mention that this study is not concerned with the
temperature dependence of the work functions. While motivated by applications
and devices which operate at elevated temperatures the non-trivial
inclusion of finite temperatures in DFT calculations is beyond the scope of
this study.

\section{Results A: External parameters}
\label{resultsA}

\begin{table*}[]
  \begin{ruledtabular}
    \caption{Work function of SrRuO$_3$ and SrTiO$_3$ with either SrO or
      (Ru,Ti)O$_2$  surface terminations. The table summarizes dependencies of the
      work function for i) different choices of the DFT functional (columns 3 \& 4
      ), ii) an unrelaxed lattice (column 5), iii) compressive (a=3.80\AA) or
      tensile (a=4.00 \AA) strained substrate (columns 6 \& 7), iv) ferromagnetic
      ground state of SrRuO$_3$ with an on-site Coulomb repulsion U=2.0eV (column
      8), v) a monolayer film; the low temperature orthorhombic structure of
      SrRuO$_3$; SrTiO$_3$ with a TiO$_2$ 2$\times$1 surface reconstruction
      (columns 9,10,11), vi) with surface oxygen vacancies in top layer/subsurface
      layer (column 12), vii) values observed in experiment}
    \label{Tableone}
    \begin{tabular}{cl|ll|l|ll|l|lll|l|l}
      \multicolumn{1}{l}{} \textbf{$\Phi$ in eV} & term.   & LDA   & GGA       & unrelaxed & \textit{a}=3.80 & \textit{a}=4.00 & U=2eV & monolayer & orthorhombic & reconstruct & Ov  & exp         \\
      \hline
      \multirow{2}{*}{SrRuO$_3$} & SrO   & 2.80       & \textbf{2.39}  & 1.30    & 2.00            & 2.55            & 2.37  & 2.60      & 2.29      & -              & 2.05/2.39   & \multirow{2}{*}{5.2 \footnotemark[1]}  \\
      & RuO$_2$          & 5.01        & \textbf{4.88} & 3.90    & 5.54               & 4.92            & 5.33     & 4,95      & 5.05      & -              & 5.03/4.91   &              \\                    
      \multirow{2}{*}{SrTiO$_3$} & SrO        & 2.52       & \textbf{1.92}  & 0.82    & 1.69            & 2.04            & -     & 2,02      & -         & -              & 2.26/1.33 & 2.4      \footnotemark[2]       \\
      & TiO$_2$        & 4.67              & \textbf{4.48}  & 3.70    & 4.47            & 4.51            & -     & 4.18      & -         & 6.18           & 3.39/3.86 & 4.6   \footnotemark[2] 
    \end{tabular}
    
  \end{ruledtabular}
  \footnotetext[1]{Fang \emph{et al.} Ref. \onlinecite{Fang:99} with unknown surface termination}
  \footnotetext[2]{Susaki \emph{et al.} Ref. \onlinecite{Susaki:prb11}}
\end{table*}

Part of our first set of calculations in which we identify key parameters that
alter the work function can be considered as DFT benchmarks. Obviously, if
the parameter in question is experimentally accessible (like e.g. substrate
strain), one can deduce potential tuning parameters of $\Phi$. 

The results we report in this section are obtained for SrRuO$_3$ and
SrTiO$_3$. Both materials are well studied and experimental data for their
work functions are available \cite{Fang:99,Susaki:prb11}. SrRuO$_3$ is a 4$d$
system and a ferromagnetic metal \cite{Koster:rmp12}; SrTiO$_3$ on the other hand is a 3$d$ non-magnetic
insulator. At this point we should make some more specific remarks about how
we deal with the calculation of work functions for the insulating SrTiO$_3$. The
difficulty for insulators is the uncertainty of the Fermi energy which needs
to be subtracted from the vacuum potential to yield $\Phi$. Instead, we decided to
consider the bottom of the conduction band as the Fermi energy due to a simple
and pragmatic argument: Our choice corresponds to the electron doped
version of the material which can be realized in experiment by La substituting Sr
\cite{Ohta:jap05}, or by Nb substituting Ti. The later technique was used in a work function study for SrTiO$_3$ by Susaki \emph{et
  al.}\cite{Susaki:prb11} and as one can see in Fig.\ref{Fig1} thei calculated
values based on our definitions are in satisfactory agreement with the experimental observations.  
Moreover, even without active doping, the (very common) occurrence of oxygen
vacancies in TMO surfaces effectively lead to the same kind of doped
electronic structure. Let us anticipate already here that our
calculations, which include such oxygen vacancies explicitly, do not capture effects
from an insulator to metal transition but rather from very small to very
large concentration of oxygen vacancies.
For these cases of slightly doped insulators, where the work function might rely
sensitively on the size of the gap between conduction and valence band, we
also make sure that DFT-GGA, which is known to underestimate gap sizes and the
Heyd-Scuseria-Ernzerhof (HSE) hybrid functionals \cite{Heyd:jcp03,
  Silva:prb07} (known to yield better results for band gaps) yield consistent
results for $\Phi$. We summarize our parameter study in
Table~\ref{Tableone}. Here, we report values for both materials and consider
either a SrO terminated or a (Ru,Ti)O$_2$ terminated (001) surface. 

The values of $\Phi$ in the first column of Table~\ref{Tableone} obtained with plain GGA 
for relaxed slabs already show an extremely important effect that we observe
in basically all calculations we have performed: Different from simple metals like
tungsten or silver, where the work function shows a surface dependence on the
order of hundreds of meV \cite{Michaelson:jap77}, the work function for perovskite oxides shows a much
more severe modulation with the choice of a specific surface, e.g. if it is AO
or BO$_2$ terminated. From our calculations we observe a difference of
$\Phi_{\rm BO_2}-\Phi_{\rm AO}=2.49$eV ($2.56$eV) for SrRuO$_3$ (SrTiO$_3$)
which prohibits clearly an approximation of $\Phi$ by a single $\Phi_{\rm
  ABO_3}$ value for oxide materials and sets the challenge for a theory/experiment
comparison: control of the sample on sub-unit cell scale seems necessary in
synthesis to support/falsify predictions from computer simulations.
If such control is not possible, the samples might have mixed termination and,
hence, display a strong sensitivity of the work function to details of the
sample preparation. Such difficulties might be one of the reasons that as of
yet there are only few experimental studies on TMO work functions\cite{Fang:99,Susaki:prb11,Greiner:afm12}. Moreover,
these complications also affect the conception of interface devices like,
e.g., TMO Schottky barriers at metal/semiconductor interface where the barrier
height is calculated with the work function of the metal \cite{Hikita:apl07}.
We will return to the discussion of termination dependent work function in the
context of building up superstructures (see section \ref{resultsC}).  

Let us turn to the comparison of GGA with LDA results. Our test cases are
actually well in line with an extensive study of Singh-Miller and
Marzari\cite{Singh-Miller:prb09}, where the functional dependence of DFT
workfunctions for metallic surfaces is discussed. The differences between GGA
and LDA can be attributed on the one hand to differences in the relaxed
structure (since GGA, e.g. generally overestimates bondlengths when compared
to experiment). On the other hand, when performed for identical lattices LDA tends
to yield always somewhat larger values than GGA. As can be seen in 
Table~\ref{Tableone} we observe total differences between $\approx
0.13-0.50$eV. While these differences are surely non negligible and one should
be aware of possible error bars. Relative values and materials trends,
however, are not affected. 

More crucial than the choice of the particular DFT functional is, however, the relaxation
of the atomic positions in the unit cell near the surface with respect to an
unrelaxed surface. Work functions calculated with unrelaxed surfaces differ partially more than $1.0$eV from the
relaxed calculations. In our calculations the surface relaxation always
increases the work function which means that the surface dipole-field increases. 
It is tempting to attribute this increase just to a surface buckling that features
an outward shift of oxygen ions at the surface (stronger for AO than for
BO$_2$ terminated surfaces). On quantitative levels a purely ionic picture is,
however, misleading since it disregards effects of relaxation of the
electronic charge involving interlayer charge transfer. The sensitivity that
we observe here indicates already the strong dependence of the work function
on microscopic details of the surface as can be seen also in the calculations
for either compressive ($a=3.80$\AA) or tensile ($a=4.00$\AA) strain which can
be achieved with growing the material on specifically chosen substrates. It
turns out that in this way modification of the work function can be achieved over a range of
up to $\approx 0.7$eV. In the SrO terminated compounds we always find a
decrease (increase) of the work function upon compressive (tensile) strain. In
the BO$_2$ terminated compounds we see a clear difference between the metallic
RuO$_2$ layer which shows an increase of the work function upon either
compressive or tensile strain, while the TiO$_2$ terminated systems is
affected by the pressure only on a very small scale compared to the other
cases.

Next we turn to the question whether a Hubbard U interaction parameter on the
B atom d-shell treated on the mean-field level has impact on $\Phi$. Such additional local potential will only
have impact in cases of partially filled shells which is why an interaction
$U=2.0$eV was taken into account only for the ruthenate calculation. We have
carried out the GGA+U calculation \cite{Dudarev:prb57} where we allowed for a
ferromagnetic symmetry broken ground state. It does not come as a big surprise
that the RuO$_2$ terminated surface is more influenced by the U on the Ru
d-shell which results in a work function enhancement by $\approx 0.4$eV while
the AO terminated surface is basically unaffected. For the general case,
however, please note that electronic interaction/correlation (approximated on
the Hartree level or beyond) might trigger phase transitions that result in a
charge redistribution, e.g. surface charge-ordered states
\cite{Hansmann:prl13} which might have significant impact on the surface
dipole field and, hence, its work function. Let us briefly point out that we
did not consider a GGA+U calculation for the band-insulating SrTiO$_3$ with a
practically empty d-shell. U would simply enlarge the gap by pushing
up empty states. Since not much is gained by such a
manual gap renormalization we state that the most reasonable step would 
rather be a GW calculations without adjustable parameters which, however, is
beyond the scope of our current study.

The following three calculations consider again more structural effects: In
monolayer setups effects of quantum confinement can alter the electronic
structure \cite{Yoshimat:sci11, zhong:prb13}. Also we remark that for most
ABO$_3$ materials, the low temperature structure is not cubic but often shows
orthorhombic distortions with tilted and rotated BO$_6$ octahedron, which,
however, do not alter $\Phi$ dramatically. Moreover, we argue that
simulations for $\Phi$ should rather consider the materials structure at the operation
temperature of the hypothetical device. Also, depending on the temperature, we take
into account that for real oxide surfaces, various surface reconstructions
exist (for example of SrTiO$_3$ \cite{Kawasaki:sci94, Koster:apl98,
  Erdman:nat02,Herger:prl98}). It is reasonable to assume that the different
structure of bonds and hence electronic densities in reconstructed surfaces
will lead to specific dipole fields and, hence, altered work functions. We
confirmed this hypothesis by studying a well-established so called
(2$\times$1) surface reconstructed phase of SrTiO$_3$ \cite{Erdman:nat02},
which can be viewed as a double TiO$_2$ layer. It turns out that the
reconstruction has a great influence on the work function which is, with a
value of $6.18$eV, much higher than the $4.48$eV of the bare TiO$_2$ surface. 

Finally, it is a well known issue for oxide surfaces that defects in the form
of oxygen vacancies should not be disregarded \cite{Nakagawa:Natm06,SusakiMgO}.
While there is a certain amount of control over oxygen vacancies in synthesis
(e.g. adjusting the oxygen pressure and annealing) the exact concentration and
distribution is generally unknown and hard to pin down. Such
defects pose a real challenge to comparing different experiments, but also experiment to
an electronic structure calculation of the oxide surfaces. The best one can do
in a calculation is to assume periodic vacancies in supercells. In our case we
assume a $2\times 2$ supercell and introduce for each case considered an
oxygen vacancy in the surface or the first subsurface layer. With this setup
we actually assume a quite high concentration of oxygen vacancies so that our
results for $\Phi$ might be considered as an upper bound of the O vacancy
effect.

As conclusion of this section stands a classification of external parameters
by means of their impact on a materials work function. The first and most
important message is that for transition metal oxide work functions the
microscopic structure of the surface electronic states/density \emph{does}
matter crucially. While our analysis underlines the challenging (but nowadays
feasible) necessity of
experimental control on the atomic scale it also tells us that a materials
 \emph{work function can be tuned}  with a number of external
 parameters. While magnetism (in the tested cases), interaction effects or
 even ``quantum confinement'' effects are not major ($\Phi$ converges rather
 quickly as a function of thickness), clean terminations and control of
 surface reconstructions is absolutely mandatory. The latter parameters can
 tune  the work function on the scale of electron volts. On a smaller scale
 ($\approx 0.5$eV) the work function can be modified, i.e. fine tuned, by
 exerting control on the oxygen defect structure and/or the choice of
 substrate. With these results in mind we will now turn to another type of
 ``control parameter'': The choice of alkali earth cation A cation and
 transition metal element B.

\section{Results B: ABO$_3$ material trends}
\label{resultsB}

As mentioned before the results in this section were obtained from setups
with a fixed in-plane lattice constant of a=3.905\AA which corresponds to
growth on a SrTiO$_3$ substrate. To disentangle trends originating in the
specific choice of cation (A)
and TM (B) from other parameters (see previous section) we consider
defect free, relaxed structures with well defined terminations in a GGA
slab calculation. The results are reported in Fig. \ref{Fig2} and
\ref{Fig3} (and corresponding data tables in appendix A).

\subsection{Termination dependence in different materials}
\begin{figure}[t]  
  \includegraphics[width=0.5\textwidth]{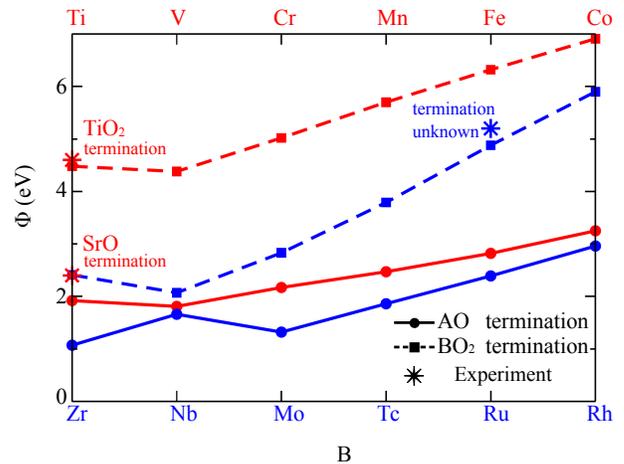}
  \caption{\label{Fig2} DFT calculated work functions of the perovskite
    ABO$_3$ series with A=Sr and \textit{B} being an element of the 3d (red)
    or 4$d$ (blue) period, and with AO (circle with solid lines)  and
    BO$_2$ (square with dashed lines) surface termination. Experimental work
    function of SrTiO$_3$ with SrO and TiO$_2$ termination by Susaki
    \emph{et al.} \onlinecite{Susaki:prb11} 
    as well as SrRuO$_3$ with unknown termination by Fang \emph{et al.}
    \onlinecite{Fang:99} are shown as stars.}
\end{figure}

\begin{figure}[t]
  \includegraphics[width=0.5\textwidth]{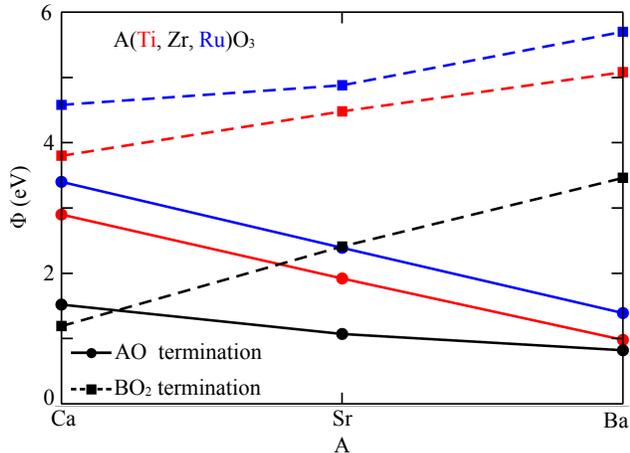}
  \caption{\label{Fig3} DFT calculated work functions of the  ATiO$_3$
    (red), AZrO$_3$ (black), and ARuO$_3$ (blue) series with A=Ca, Sr and
    Ba, and with AO (circle with solid lines)  and BO$_2$ (square with dashed
    lines) surface termination.}
\end{figure}

Overall we find as a first remarkable fact a confirmation of the
crucial dependence of $\Phi$ on the termination, see Fig. \ref{Fig2} and
\ref{Fig3}. Except for a single case (CaZrO$_3$) the work function of the AO
terminated surface is smaller than the BO$_2$ terminated one.
It turns out that this observed materials dependence hints towards a new twist to the
interpretation of the termination sensitivity: We remind ourselves that the
surface dependence of the work function originates in the dipole field created
by polarization of the electronic charge and shifts of atoms/ions close to the
specific surface. Two effects which are obviously coupled in our calculations which
include a self consistent lattice relaxation. While this interplay is quite
involved and cannot be disentangled easily, we observe a clear
correlation between the work function behavior and the 
\emph{electronegativity $\chi$} of the cation and transition metal elements \cite{Allen:jacs89}. 

The concept of electronegativity is usually used in order to estimate the
character of an ionic bond in a binary compound. Taking the difference of the
two elements electronegativity allows for classification of the bond into
either ionic or covalent. So \emph{$\chi$ reflects the ability of an atom to attract
electron density}. Keeping this in mind it comes not as a big surprise that for
simple metals the compounds work function is linked to the elements
electronegativity\cite{Trasatti}. Our results and analysis show that, remarkably we can
still use the electronegativity concept in our ternary compounds.
Following the Allen scheme\cite{Allen:jacs89} of electronegativity we find a
monotonous increase in $\chi_B$ from Ti ($\chi_{\rm Ti}=1.38$) to Co
($\chi_{\rm Co}=1.84$) for the 3d series, a monotonous increase from
Zr($\chi_{\rm Zr}=1.32$) to Rh ($\chi_{\rm Rh}=1.56$) for the 4d series, and
for the cations we have $\chi_{\rm Ca}=1.03$, $\chi_{\rm Sr}=0.96$, and
$\chi_{\rm Ba}=0.881$ \footnote{For our qualitative discussion the specific
  type of the electronegativity (e.g. Pauling, Muliken, Allan, etc.) is not
  really important since materials trends are very similar for all schemes.}
Oxides are typically considered as very ionic due to the high electronegativity
of oxygen $\chi_{\rm O}=3.61$. 

For our materials it turns out to be useful to introduce the idea of a layer
electronegativity $\chi_{AO}$ and $\chi_{BO_2}$. In all considered compounds
$\chi$ of cation A is smaller than that of transition metal B so that on the
one hand A-O bonds can be considered as more ionic than B-O bonds but also
that the average AO electronegativity is smaller than that of
the BO$_2$ layers $\chi_{AO}<\chi_{BO_2}$. This explains the general tendency
of smaller AO work functions $\Phi_{AO}<\Phi_{BO_2}$ which we already reported.
It turns out that with these rough estimates many of the following materials
trends can be explained.

\subsection{Control via transition metal B for A=Sr}
We will now discuss tuning of the work function with choice of the transition
metal element B in more detail. As shown in Fig.~\ref{Fig2} we have performed
calculations for a series of 3$d$ and 4$d$ transition metal compounds. The overall
variation of the work function is a remarkable $6$eV ($\approx \Phi^{\rm
  BO_2}_{\text{SrCoO}_3}-\Phi^{\rm AO}_{\text{SrZrO}_3}$ ). Continuing the line of
argument from above, the trends we observe can be explained by comparison of
$\chi_B$ of the transition metal element (or the effective BO$_2$
electronegativity $\chi_{BO_2}$). It is no surprise that the
trend is weaker in the case of AO termination since increased $\chi_{BO_2}$
in the subsurface layer has less effect on $\Phi$. The increasing $\Phi$ within
each series (3$d$,4$d$) reflects precisely the increase of $\chi_{BO_2}$ within
the period and so does the decrease of $\Phi_{3d} < \Phi_{4d}$ reflect
$\chi_{3d} < \chi_{4d}$.  A closer look at the numbers shows indeed quite low
work functions for AO termination throughout the series and an almost
monotonous increase from $\Phi_{\text{SrTiO}_3}\approx 1.9$eV (3$d$ series)
and $\Phi_{\text{SrZrO}_3}\approx 1.1$eV (4$d$ series) \footnote{Please note
  that SrTiO$_3$ as well as SrZrO$_3$ are band insulators such that we took
  the position of the lowest conduction band as reference energy for the work function.} 
to $\Phi_{\text{SrCoO}_3}\approx 3.2$eV and
$\Phi_{\text{SrRhO}_3}\approx 3.0$eV with the only exception of $\Phi_{\text{SrNbO}_3}$
being slightly higher than $\Phi_{\text{SrMoO}_3}$.
We find the same trend in the BO$_2$ terminated surfaces of these materials
though at values for $\Phi$ which are roughly larger by a factor of 2.

Before we continue our discussion for cation controlled tuning of $\Phi$ let
us compare the results to the few experimentally available data points (shown
as ``stars'' in Fig.~\ref{Fig2}). For
SrTiO$_3$ both values for TiO$_2$ or SrO terminated surfaces \cite{Susaki:prb11} are in very
satisfactory agreement with our calculations. (The comparison shows further
that due to the very likely presence of oxygen vacancies it is a reasonable
ansatz to take the energy of the lowest (electron doped) conduction band  as
reference energy for $\Phi$). The only other data point is that of
SrRuO$_3$ for which, however, the precise termination was undetermined\cite{Fang:99}. In
fact one is tempted to conclude by comparison to our results that the termination was
most likely a RuO$_2$ dominated one. Yet, one has to be careful since rough
surfaces are most probably determined not only by a mixture of AO and BO$_2$
domains but also by polarizability of defects like domain walls etc.. We use
this observation to emphasize once more the importance of microscopic control
of the surface structure if comparison or predictions of calculations like
ours should be considered. Moreover, if oxide heterostructures are used in
devices like Schottky barriers \cite{Hikita:apl07,
  Minohara:prb10,Yajima:natc15} or for organic electronics
\cite{Greiner:natm12, Greiner:NPGAsia13} the oxide work function is often used
as for simple metals or semiconductors. We emphasize once more, however, that this is very
dangerous since \emph{there is no such thing as a single valued work function for
  an oxide material}.  

\subsection{Control via cation A}
We now turn to control of the work function with the choice of the alkaline
earth cation A. In Fig. \ref{Fig3} we report results obtained for nine
selected compounds (Ca,Sr,Ba)(Ti,Zr,Ru)O$_3$ to study trends with the cation
choice. As a first observation we state that the overall dependence of $\Phi$ is
quite large and that the cation choice is apparently a promising tuning
parameter. The changes in the AO terminated surfaces are more sizable than before and
follow the trend of electronegativity of the cation: Ba has the smallest
$\chi$, and hence the smallest work functions. 
However, the trend in BO$_2$ terminated surfaces with cation A is not as
easily explained! While $\chi_A$ decreases from Ca to Ba, the work function for
BO$_2$ terminations actually increases. This points towards an aspect which is
not taken into account by our simple electronegativity argument: \emph{Charge
transfer} between AO and BO$_2$ layers. One possible explanation is that more
ionic AO layers lead to an increased charge transfer to the terminating BO$_2$ layer
which then increases the dipole field and, hence, the work function. On the
other hand we cannot underline this hypothesis with evidence and remain with
reporting the observed trend.

\subsection{Promising materials}
At the end of our materials study we can confirm experimental strategies with our calculations
 that have been established on an empirical basis in the past. The
electronegativities $\chi_A$ and $\chi_B$ might be used (keeping the
limitations of the estimate in mind) as a rough guidance to select promising
materials. Our considered materials cover a wide range of work functions 
reaching from $6.91$eV (CoO$_2$ terminated SrCoO$_3$) down to $0.82$eV 
(BaO terminated BaZrO$_3$). For electron emitting devices and, more specifically, 
emitter and collector materials in thermoelectronic setups one needs low work functions. 
These are found in the compounds of the early elements in the 3$d$ and 4$d$ 
series: First of all we point out SrO terminated SrMoO$_3$ ($\Phi^{\rm
  SrO}_{\rm SrMoO_3}$)which is the material with the highest conductivity in
our 4$d$ series. It can be grown in thin films by pulsed laser deposition\cite{Radetinac:ape10}
and would be a good candidate, e.g. for the electrode material in
thermoelectronic devices. Another very interesting compound we would like to
highlight is CaZrO$_3$. It is remarkable due to the fact that the work
function is at the same time quite low \emph{and} rather similar for both
terminations (see Fig.\ref{Fig3}). This means, that also less clean
(heterogeneous) CaZrO$_3$ (001) surfaces, which are much less
tedious/expensive to synthesize, will be good electrode materials.  Third, our
results for BaBO$_3$ are in agreement with the well known observation that
coverage with BaO will lower a materials work function \emph{if AO terminated
  systems are concerned}. This last conclusion leads us to our third and last
section where we discuss how to tune the work function by building
heterostructures. 

\begin{figure}[h,t]
  \includegraphics[width=0.5\textwidth]{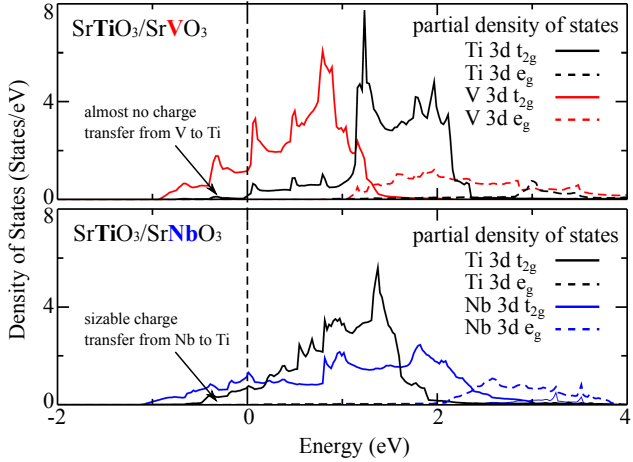}
  \caption{\label{Fig4} DFT calculated density of states for
    SrVO$_3$ capped with one unit cell
    of SrTiO$_3$ (upper panel) and SrNbO$_3$ capped with
    one unit cell of SrTiO$_3$ (lower
    panel). We show only the partial density of states for V, Nb, and Ti
    3d-states.$t_{2g}$ and $e_g$ states are indicated by solid and dashed lines respectively. The figure shows a clear material dependence of the charge
    transfer between base and capping material.}
\end{figure}

\section{Results C: Oxide heterostructures}
\label{resultsC}
In the previous section we have already made several observations how (and
gave arguments why) changes in surface layer and charge
transfer between layers close to the surface affect the work function.
In the third and final part of our study we turn to even more drastic
surface manipulation: instead of just choosing one of the materials lattice
planes to be the surface layer we build the surface actively by combining
different compounds.

To this end we model symmetric A'B'O$_3$/ABO$_3$ heterostructures by adding
A'B'O$_3$ thin films (grown in the 001 direction) with varying number of unit
cells N on both sides of the ABO$_3$ slab (i.e. capping) with either AO or
BO$_2$ termination. To be more specific, an AO terminated material capped with
A'B'O$_3$ will have an A'O interface with vacuum while a BO$_2$ terminated one
will have a B'O$_2$ interface with vacuum.   

\begin{table}[h,t]
  \begin{ruledtabular}
    \caption{Work function of SrVO$_3$, SrNbO$_3$ and SrRuO$_3$ capped by SrTiO$_3$ thin films. The thickness of SrTiO$_3$ N is varied from 0 to 2}
    \label{Tabletwo}
    \begin{tabular}{lllllll}
      SrTiO$_{3}$ & \multicolumn{2}{l}{SrVO$_3$} & \multicolumn{2}{l}{SrNbO$_3$} & \multicolumn{2}{l}{SrRuO$_3$} \\
      
      & SrO         & VO$_2$         & SrO          & NbO$_2$        & SrO          & RuO$_2$        \\
      \hline
      N=0                        & 1.81        & 4.38        & 1.66         & 2.07        & 2.39         & 4.88        \\
      N=1                        & 1.88        & 4.61        & \textbf{1.18}         & 3.41        & 2.17         & 4.72        \\
      N=2                        & 1.87        & 4.77        & 1.29         & 3.82        & 2.13         & 4.87       
    \end{tabular}
  \end{ruledtabular}
\end{table}
\begin{figure}[h,t]
  \includegraphics[width=0.5\textwidth]{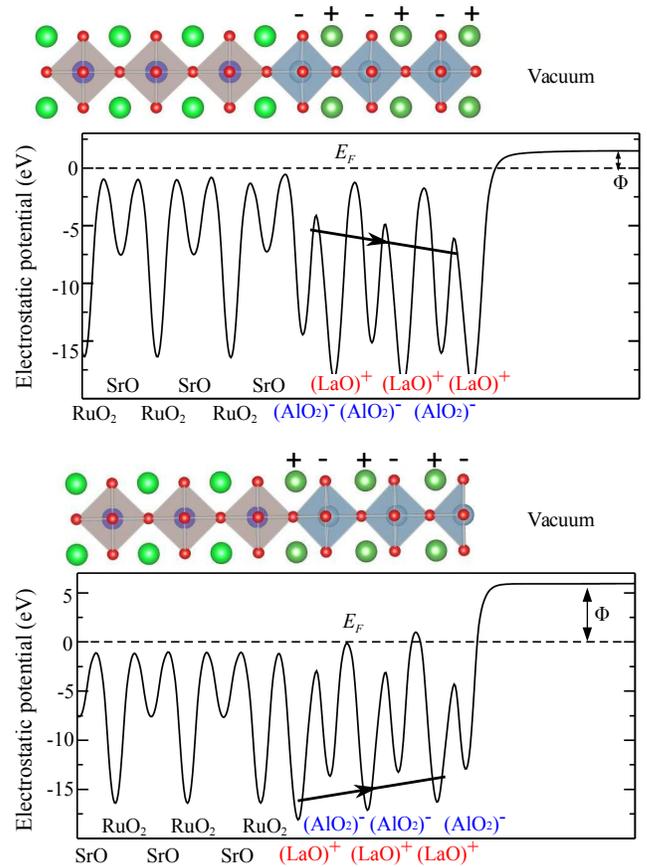}
  \caption{\label{Fig5} Plane averaged electrostatic potential for a
    LaAlO$_3$/SrRuO$_3$ heterostructure calculated with DFT: three layers of
    LaAlO$_3$ grown on a SrRuO$_{3}$ substrate with AO termination (upper
    panel) or BO$_2$ termination (lower panel). The internal field of the
    polar layers tunes the work function depending on its direction.}
\end{figure}

\subsection{Non-polar capping:}
In table \ref{Tabletwo} we present results for capping three example materials with
SrTiO$_3$ of varying thickness of either one or two unit cells (as well as the
reference N=0 for the uncapped material). SrTiO$_3$ is a charge neutral and
non-polar band insulator. The results depend strongly on the specific
case. The most severe change is found in SrNbO$_3$ capped with one unit cell
of SrTiO$_3$. Here the work function is reduced for the SrO termination from
$1.66$eV to $1.18$eV while the NbO$_2$ terminated (i.e. TiO$_2$ as final
layer) compound experiences an increase of the work function from $2.07$eV to
$3.41$eV. By comparison changes in SrVO$_3$ or SrRuO$_3$ are less
pronounced. The reason for this different behavior is found when we study the
charge transfer between base and capping material. In Fig.~\ref{Fig4} we show
the partial density of states for SrTiO$_3$/SrVO$_3$ (upper panel) and
SrTiO$_3$/SrNbO$_3$ (lower panel).  

The local potential of Ti d orbitals in SrTiO$_3$ turns out to be comparable
to the Nb d potential in SrNbO$_3$. V d orbitals in SrVO$_3$, however, reside at
a much lower energy \cite{zhong:epl12}. As a consequence we encounter a
substantial charge transfer in SrTiO$_3$/SrNbO$_3$ (as seen in
Fig.\ref{Fig4}), while basically no charge transfer occurs in SrTiO$_3$/SrVO$_3$. 
As one can from the lower panel the Ti d-states in the SrTiO$_3$/SrNbO$_3$
system are heavily electron doped by the base material while this is not the
case for the vanadate case.  Such doping has severe consequences not only by
shifting the Fermi level into the Ti d-states but also by actually charging
the capping layer which alters the surface dipole field. It is interesting how
such complex interplay of lattice, orbital and charge degrees of freedom
eventually leads to a work function for the AO terminated surface which is
lower than that of the parent compounds SrTiO$_3$ or SrNbO$_3$. This case
confirms, as a proof of principle, the potential of tuning $\Phi$ by
heterostructuring oxide materials.  

One should remark that in the example we have just discussed the capping
material was an insulator. If we choose A'B'O$_3$ to be a metal, we see that the
work function is dominated by the A'B'O$_3$ thin films (so that the capped
ABO$_3$ material is almost entirely irrelevant). This is simply due to the
screening properties of the metallic capping compound and we could only study
the strain and quantum confinement effect in A'B'O$_3$.

\begin{figure}[t]
  \includegraphics[width=0.5\textwidth]{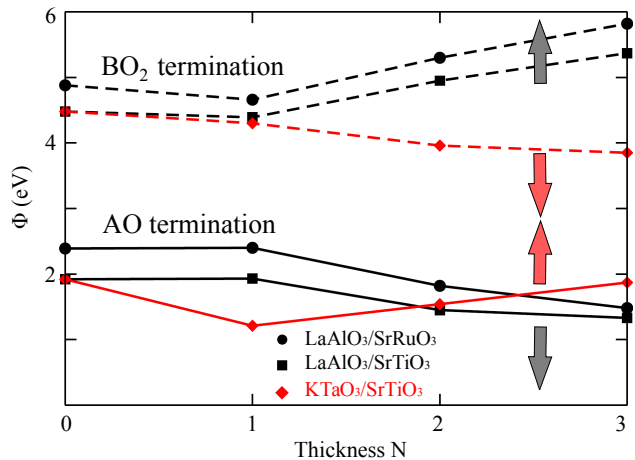}
  \caption{\label{Fig6} Work function plotted versus thickness of polar
    capping material. As already indicated in Fig.~\ref{Fig5} the additional
    field produced by polar layers can be used to tune the work function in
    either way. Here we show how either termination can be tuned up or down
    by choosing the appropriate capping material LaAlO$_3$ or KTaO$_3$ which
    have opposite effects due to their opposite polarity.}
\end{figure}

\subsection{Inducing intrinsic fields by polar capping}
When A'B'O$_3$ is a polar insulator, such as LaAlO$_3$, which can be viewed as
an alternating stack of positively charged (LaO)$^+$ layers and negatively
charged AlO$_{2}^{-}$ layers.  When it is grown on a non-polar compound like
SrRuO$_3$ or SrTiO$_3$ (charge neutral (SrO)$^0$ and (Ru,Ti)O$_{2}^{0}$
layers), A possible hypothesis that was studied in the past
\cite{Susaki:prb11} is that thin films of a polar material (e.g. LaAlO$_3$)
play the role of a parallel capacitor that introduces an internal electric
field pointing from surface, e.g. LaO, layer to the interface of
e.g. SrO/AlO2. The resulting potential drop should lead, as the polar capping
material thickness increases, to a decrease of the work function. We sketch
this scenario in Fig.~\ref{Fig5}: the result for LaAlO$_3$ (KTaO$_3$) capping
would be a decrease (increase) of the work function for the AO and an increase
(decrease) for the BO$_2$ terminated case. We performed calculations for
SrRuO$_3$ or SrTiO$_3$ for LaAlO$_3$ and KTaO$_3$ capping of different
thickness and present the results in Fig.~\ref{Fig6}. Let us first remark that
the comparison between N=0 and N=1 is always delicate since we 
introduce not only the additional field but also the surface relaxation changes most
from the uncapped to the N=1 case. The data from N=1 to N=3 shows that the DFT
calculation confirms the simple hypothesis for all checked cases: LAO capped
AO surface layers show decreasing $\Phi$ with increasing N while LAO capped
BO$_2$ terminated cases show an increase. The opposite is true for capping
with KTaO$_3$. In principle this observation is encouraging. However, these
observations are in contradiction to experiments performed by Susaki et
al.~\cite{Susaki:prb11}. In their article the authors already mention the
discrepancy of the simple ``additional internal field'' picture with their
measured data: Instead of a work function increase they found a remarkable
decrease in TiO$_2$ terminated SrTiO$_3$ capped with LaAlO$_3$.    

In summary we report that DFT results agree with the simple picture of
superimposed potentials but not with experiment. A reasonable explanation for
the discrepancy might be found in defect structures not taken into account by
our calculations. In section~\ref{resultsA} we have seen that oxygen vacancies
can alter the work function on a significant scale so that a disregard of
likely defects in the LaAlO$_3$ capped systems is not justified.  

\section{Conclusion:}
In conclusion we have presented a systematic work function study for transition
metal compounds which allowed us to classify the sensitivity of the work function
to parameters of DFT calculations as well as to experimentally accessible
 conditions like oxygen vacancies, substrate strain, and heterostructuring.
We were able to conclude, on general grounds, that the work function concept is
more complex in oxide materials than in simple metals and that microscopic
details at the materials surface matter crucially. We emphasize that tuning oxide work functions
(predictably) in the experimental lab generally requires synthesis control on
the atomic scale. This challenge comes, however, with the prize that oxide
work functions are indeed highly tunable and, depending on the
parameter can be manipulated on different energy scales. While the choice of
the material and the choice of the terminating layer tunes oxide work
functions over several eV, substrate induced strain or a suitable capping
material can fine tune the work function on the sub-eV scale to the desired
value. We have also uncovered that, like for simple metals, there is a link
between the observed work function $\Phi$ and electronegativities $\chi_A$ and $\chi_B$ of
the elements in the compound which was helpful in explaining the observed
materials trends and might prove to be also useful for extrapolating our
results in other directions. For the manipulations of $\Phi$ with polar
capping layers we conclude from our theory/experiment comparison with a
warning that for real materials an oversimplified electrostatic picture is
highly doubtful.

Let us finally remark that one of the main intentions of this manuscript is to form a
fix point for future studies. The ``phase space'' of materials and tuning
parameters is infinitely large so that any systematic search needs a well
established base. We hope that also experimental colleagues can help to judge
the quality of the many data points we predicted for materials that have not yet
been measured in order to establish such a base.

\section{Acknowledgments}
We thank J. Mannhart, I. Rastegar, G. Giovannetti, N. Spaldin, and L. Giordano for motivation and useful discussions.

\appendix

\section{$\Phi$ Tables}
\begin{table}[h]
\begin{ruledtabular}
\caption{Data of Fig.~\ref{Fig2}: Work functions of SrBO$_{3}$ (B=3$d$, Ti-Co; 4$d$: Zr-Rh) with SrO and BO$_2$ surface terminations}
\label{Tablethree}
\begin{tabular}{lllllll}
3$d$ SrBO$_{3}$   & Ti & V &  Cr & Mn &Fe &Co\\
                         \hline
AO &1.92     & 1.81        & 2.17        & 2.47        &2.82         &3.25       \\
BO$_2$ &4.48        & 4.38        &5.02        & 5.70        &6.32         & 6.91        \\
\hline
4$d$ SrBO$_{3}$ & Zr & Nb & Mo & Tc & Ru &Rh \\
AO &1.07     &1.66        &1.32       & 1.86        &2.39        &2.96      \\
BO$_2$ &2.41       & 2.07        &2.83        &3.79       &4.88        & 5.90
\end{tabular}
\end{ruledtabular}
\end{table}

\begin{table}[h]
\begin{ruledtabular}
\caption{Data of Fig.~\ref{Fig3}: Work functions of ABO$_{3}$ (A=Ca, Sr, Ba; B=Ti, Zr, Ru) with AO and BO$_2$ surface terminations}
\label{Tablefour}
\begin{tabular}{lllllll}
  & \multicolumn{2}{l}{ATiO$_3$} & \multicolumn{2}{l}{AZrO$_3$} & \multicolumn{2}{l}{ARuO$_3$} \\
                         & AO         & TiO$_2$         & AO          & ZrO$_2$        & AO          & RuO$_2$        \\
                         \hline
A=Ca                      &2.90        & 3.80    &1.52         & 1.19       & 3.40        &4.58             \\
A=Sr                       & 1.92        & 4.48    & 1.07         &2.41         & 2.39         &4.88           \\
A=Ba                       & 0.98        & 5.08      & 0.82         & 3.46       & 1.39         & 5.70        
\end{tabular}
\end{ruledtabular}
\end{table}

\clearpage

\end{document}